\documentclass[aps,twocolumn,groupedaddress,showpacs]{revtex4}

\usepackage{epsfig,amsbsy,bm}
\newcommand{\eref}[1] {(\ref{#1})}
\newcommand{\Eref}[1] {Eq.~(\ref{#1})}
\newcommand{\Fref}[1] {Fig. \ref{#1}}

\newcommand{\be}{\begin{equation}}
\newcommand{\ee}{\end{equation}}

\renewcommand{\bbox}{\boldsymbol }

\newcommand{\np}{\newpage}
\newcommand{\bt}{\begin{tabular}}
\newcommand{\et}{\end{tabular}}
\newcommand{\bp}{\begin{minipage}}
\newcommand{\ep}{\end{minipage}}

\newcommand{\br}{\begin{eqnarray*}}
\newcommand{\er}{\end{eqnarray*}}
\newcommand{\ba}{\begin{eqnarray}}
\newcommand{\ea}{\end{eqnarray}}

\renewcommand{\k}{{\bbox k}} 

  \newcommand{\p}{{\bbox p}}

\newcommand{\isum}%
{\mathop{\hbox{$\displaystyle\sum\kern-13.2pt\int\kern1.5pt$}}}

\begin{document}

\bibliographystyle{apsrev}

\title {Extraction of  attosecond time delay using the soft photon
  approximation.}

\author{ I.A.Ivanov\footnote[1]{Corresponding author: igor.ivanov@anu.edu.au}
}
\author{A.S.Kheifets}

\affiliation
{Research School of Physical Sciences,
The Australian National University,
Canberra ACT 0200, Australia
}

\date{\today}

\begin{abstract}
  We use the soft photon approximation to extract the Wigner time
  delay from atomic two-color photoionization experiments. Unlike the
  strong field approximation, the present method does not require
  introduction of the Coulomb-laser coupling corrections and enables
  one to extract the Wigner time delay directly from attosecond time
  delay measurements.
\end{abstract}


\maketitle

\section{Introduction}

The concept of time delay was developed in formal scattering theory
by Wigner \cite{delay2} and his contemporaries (see Ref.~\cite{delay3}
for a comprehensive review).  It is a quantity related to the phase of
the complex scattering amplitude which provides an insight into
development of the scattering process in time.  In recent years, this
idea has made a dramatic comeback when it was realized that the time
delay can be measured experimentally in photoionization processes.
This has led to many interesting and not yet fully understood results
such as observation of a considerable time delay between
photoelectrons emitted from the $2s$ and $2p$ sub-shells in neon
\cite{delay0}, or an experimental determination of the tunneling time
in an ionization event \cite{angstr}.

The timing information in photoionization process is extracted
experimentally by applying an ionizing XUV pulse (the pump pulse)
followed by an infrared (IR) probe pulse.  In the attosecond streaking
experiments, the time delay between the pump and probe pulses is
mapped onto the kinetic energy of the photoelectron in the form of a
spectrogram. In such experiments, duration of the probe pulse may be
several optical cycles of the IR field \cite{delay0}.  Alternatively,
one may use the so-called RABITT (Reconstruction of Attosecond Bursts
by Ionization of Two-photon Transitions) technique \cite{rabbit} which
employs a monochromatic IR probe. In this technique, the pump-probe
delay is mapped onto the phase of the sideband oscillations caused by
interference of alternative two-photon ionization processes. A
detailed description of these techniques can be found in \cite{dal2}.

To extract the Wigner time delay related to the XUV photoionization,
one has to take into account the effect of the probe IR field on the
system under investigation. In the RABITT experiments with
monochromatic probes, the IR field is typically weak, which allows the
perturbation theory treatment \cite{dal1,dal3}. In the attosecond
streaking approach, where the IR probe intensity is typically in the
range $10^{11}-10^{12}$ W/cm$^2$, the non-perturbative treatment is
called for. In the first interpretation of the attosecond streaking
experiment \cite{camera}, the well-known classical equation was
invoked:
\be
\p_f(t) = \p_0 - {\bm A}^{\rm IR}(t) \ ,
\label{str}
\ee
relating the unperturbed asymptotic momentum of the photoelectron
$\p_0$ and the final momentum $\p_f(t)$ for emission at time $t$
in the presence of an IR field ${\bbox A}^{\rm IR}$. This implies that
the interaction of the photoionization with the ionic core is
neglected.  To account for the corrections due to this interactions
and distortion of the initial atomic state by the IR field (the
so-called Coulomb-laser coupling) , the further refinement of this
model has been developed \cite{pap3,clc,paz5}.

Below we present an alternative procedure of extraction of the time
delay from the experimentally observable photoionization
cross-sections. This procedure introduces an accurate description of
the IR field influence from the outset and no further corrections are
needed.

\section{Theory and computational details}

The procedure is based on the so-called {\em soft photon
  approximation} \cite{soft1}. Under condition of the IR photon
frequency being small in comparison with the photoelectron energy, this
approximation has been shown to reproduce quite accurately the
angle-integrated cross sections of the process of two colour
ionization by the XUV and IR fields \cite{soft2}.  To extract timing
information, one has to know the phase or, rather, the energy
derivative of the phase of the amplitude of the ionization process. It
is unclear whether the soft photon approximation can cope with this
problem.  Below, we address this question.

We consider a typical configuration of the XUV and IR fields used in
the attosecond streaking  experiments.  The time dependence of the
electric field of the IR pulse is
\be
{\cal E}^{\rm IR}(t)={\cal E}_0^{\rm IR} \sin{\Omega t} 
\ ,
\label{ir}
\ee
with the base frequency $\Omega=0.057$ a.u. ( photon energy of 1.55
eV) and the peak field strength $E_0^{\rm IR}=0.004$ a.u.  (intensity
of $5.6\times 10^{11}$ W/cm$^2$). The IR field is present on the interval
of time $(0,T_1)$, where  $T_1=2\pi/\Omega=2.7$~fs is an optical cycle
corresponding to the IR frequency $\Omega$.

The XUV pulse is present on the time interval $\Delta-4T,\Delta+4T$,
where $T=2\pi/\omega$ is an optical cycle of the XUV pulse. Parameter
$\Delta$, therefore, characterizes the relative shift between
beginning of the IR pulse and arrival of the center of the XUV pulse.
On this interval the XUV field time-dependence is
\be
{\cal E}^{\rm XUV}(t)={\cal E}_0^{\rm XUV} f(t')\cos{\omega t'}
\ ,
\label{xuv}
\ee
where $t'=t-\Delta$, and we use a cosine squared envelope function
$\displaystyle f(t')=\cos^2(\omega t'/ 16)$. The XUV field strength is
${\cal E}_0^{\rm XUV}=0.01$ a.u. (intensity of $3.5\times 10^{12}$
W/cm$^2$). Both pulses are assumed linearly polarized along the
$z$-axis.  As a target system, we consider  the Ne atom described by
 a localized model potential \cite{oep} within the single active electron (SAE)
approximation.

The amplitude of the photoionization process can be defined as
\be f(\k)= \lim_{{t\to\infty\atop \tau\to -\infty}} 
e^{i(E(k)t-E_0\tau)}\langle\Psi_{\k}^-|\hat
U(t,\tau)\phi\rangle\ ,
\label{am1} \ee
where $ \Psi_{\k}^-$ is the (ingoing) scattering wave function
describing the photoelectron with the kinetic energy $E(k)$, $\hat
U(t,-\infty)$ is the evolution operator propagating the system in
presence of the IR and XUV fields, $\phi$ is the initial atomic
state and $E_0$ is its energy.
For a relatively weak XUV field strength, the
photoionization amplitude in presence of the XUV pulse alone is given
by the well-known perturbative formula:
\be f^{\rm XUV}({\k})=-i\int_{-\infty}^{\infty} \langle
\Psi_{\k}^-|\hat H_{\rm int}^{\rm XUV}(t)|\Psi_0\rangle
e^{i(E(k)-E_0)t}\ dt \
\label{amp} \ee
Expression for the evolution operator applicable for a weak XUV
field can be obtained from the Dyson equation:
\be \hat U(t,\tau)= \hat U_0(t,\tau) -i \int_{-\infty}^t \hat U_0(t,x)
H_{\rm int}^{\rm XUV}(x) \hat U_0(x,\tau)\ dx,
\label{ev} 
\ee 
where $\hat U_0(t,\tau)$ is the evolution operator for the atom in
presence of the IR field only.  In the following, we adopt the
Coulomb-Volkov approximation (CVA) \cite{cva1,cva2}. In this
approximation,  the action of the evolution
operator $U_0(\tau,t)$ on the scattering state $\Psi_{\k}^-$ of the
atom is expressed as 
\be
 \hat U_0(\tau,t)\Psi_{\k}^- =
\Psi_{\k}^-\exp{\bigl(-{i\over 2}\int_t^{\tau}(\k+{\bm A}^{\rm IR}(x))^2
  dx\bigr) }
\ ,
\label{cvaa}
\ee
where $ \bm{A}^{\rm IR}(t)=-\int_0^t {\cal E}^{\rm IR}(x)\
dx$ is the vector potential of the IR field. We shall also make an
assumption that the IR field perturbs the initial (ground) state only
slightly. 
So we can write 
$\hat U_0(x,\tau)\phi=e^{-iE_0(x-\tau)}\phi $.

We shall consider below emission of the photoelectron in the $z$-
direction which is parallel to the polarization vectors of both the IR
and XUV fields.
By substituting \Eref{ev} into \eref{amp}, using the CVA, expanding
exponential introduced by the CVA as a Fourier series, and utilizing
the perturbative equation \Eref{amp} for the photoionization amplitude
in presence of the XUV field only, we obtain the following expression:
\be
f(k_z)=e^{i(y-E(k))T_1}\sum\limits_{m=-\infty}^{\infty}J_m\left({y\over
\Omega}\right) f^{\rm XUV}(k_z^{(m)})\ ,
\label{ampf} 
\ee 
where $y={\cal E}^0_{\rm IR}k_z/\Omega$,
$k_z^{(m)}=\sqrt{k^2-2y+2m\Omega}$ and $J_m$ is a Bessel function.
Terms with different $m$ in \Eref{ampf} describe processes with
participation of $m$ IR photons.

By using \Eref{ampf} for various delays $\Delta$ between the IR and
XUV fields, we can obtain a set of relations between the amplitudes
$f(k_z,\Delta)$ and the amplitudes $f^{\rm XUV}(k_z,\Delta)$ of the
photo-ionization driven by the XUV field alone. Here we introduced the
explicit dependence of the photoionization amplitudes on $\Delta$ for
convenience of notations.  The perturbative expression \eref{amp}
allows us to express $f^{\rm XUV}(k_z,\Delta)$ in terms of the
'reference' amplitude $f^{\rm XUV}(k_z,0)$ as $f^{\rm
  XUV}(k_z,\Delta)=e^{i(E(k)-E_0)\Delta}f^{\rm XUV}(k_z,0)$.

Our goal is to determine the phase, or rather the phase derivative, of the
reference amplitude with respect to the electron momentum, since the
quantity of interest for us, the time delay $\tau_0$ can be expressed as
\cite{circ6}:
\be \tau_0= {1\over k_z}{\rm Im} \left( {\partial f^{\rm
XUV}(k_z,0)\over \partial \k_z}\right) \ .
\label{td} 
\ee
Here the derivative is to be taken at the point $k_z$ satisfying the
energy conservation $E_0+\omega=k_z^2/2$, $E_0$ being the energy of
the initial atomic state.  By using this equation and \Eref{ampf}, it
is not difficult to devise a procedure allowing to obtain information
about the phase of the reference amplitude for the process of
photoionization by the XUV field from the experimentally measurable
cross-sections of the photoionization process in presence of both the
XUV and IR fields. Before describing implementation of such a
procedure, we have to ascertain first that \Eref{ampf} is accurate
enough.

\section{Numerical results}

To this end, we solve the time dependent Schr\"odinger equation (TDSE)
for the Ne atom described by means of the model SAE potential
\cite{oep} in presence of the XUV and IR fields given by \Eref{ir} and
\eref{xuv}.  We employ the procedure allowing us to solve numerically
a 3D TDSE which is described in details in \cite{dstrong,circ6}. By
projecting the solution of the TDSE on the scattering state
$\Psi_{\k}^-$ of the Ne atom, as prescribed by \Eref{am1}, we obtain
the photoionization amplitude $f(\k)$ in presence of both the XUV and
IR fields. A separate calculation of atomic evolution in presence of
the XUV pulse alone described by \eref{xuv} with $\Delta=0$ gives us a
'reference' amplitude $f^{\rm XUV}(k_z,0)$. By using the relation
connecting $f^{\rm XUV}(k_z,\Delta)$ and $f^{\rm XUV}(k_z,0)$ and
\Eref{ampf}, we can compute values of $f(k_z,\Delta)$, which is the
amplitude of the two-colour ionization for different values of the
delay $\Delta$ between the XUV and IR pulses, and compare them with
the {\it ab initio} values of $f(k_z,\Delta)$ provided by the TDSE
calculation. Such a comparison is shown in Figs.~\ref{fig1}, \ref{fig5}
and \ref{fig2}
below.  The data were obtained retaining the terms with
$|m|\le 5$  in \Eref{ampf}.

\begin{figure}[h]
  \begin{center}
    \begin{tabular}{cc}
      \resizebox{60mm}{!}{\includegraphics{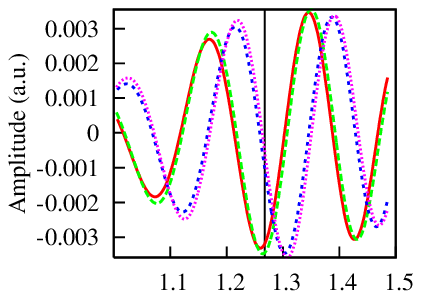}} \\
      \resizebox{60mm}{!}{\includegraphics{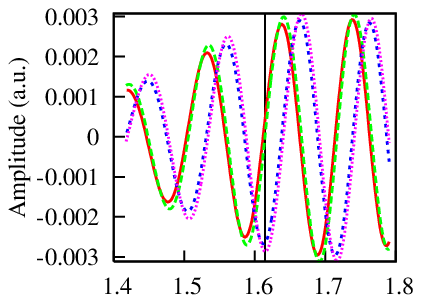}} \\
      \resizebox{60mm}{!}{\includegraphics{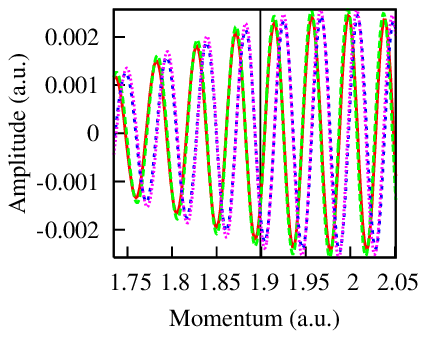}}
    \end{tabular}
\caption{(Color online) Ionization from the $2s$ state of a Ne atom.
${\rm Re} f(k_z,\Delta)$ computed
using \Eref{ampf}, solid (red) line, and TDSE calculation (green) dash.
${\rm Im} f(k_z,\Delta)$ given by 
\Eref{ampf}, (blue) short dash and TDSE (magenta) dots.
XUV photon energies (top to bottom) are $\omega=68$, 81.6, and 95~eV.
Delays $\Delta$ are (top to bottom) $0.2T$, $0.3T$, and $0.7T$, 
where $T$ is 
an optical cycle of the XUV pulse.
Vertical solid line corresponds to the momentum $k_z$ for which
$E_0+\omega=k_z^2/2$.}
\label{fig1}
\end{center}
\end{figure}

\begin{figure}[h]
  \begin{center}
    \begin{tabular}{cc}
      \resizebox{60mm}{!}{\includegraphics{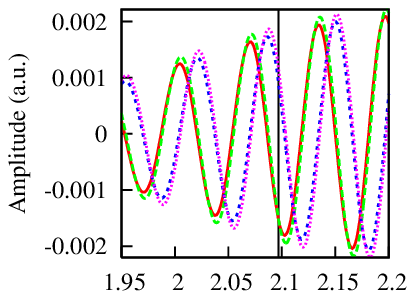}} \\
      \resizebox{60mm}{!}{\includegraphics{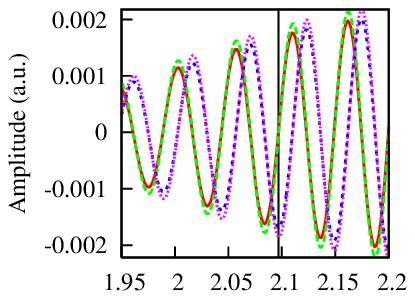}} \\
      \resizebox{60mm}{!}{\includegraphics{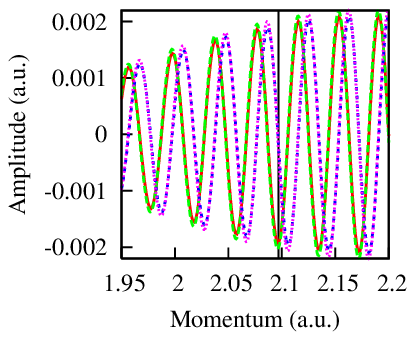}}
    \end{tabular}
\caption{(Color online) Ionization from the $2s$ state of Ne atom.
${\rm Re} f(k_z,\Delta)$ computed
using \Eref{ampf}, solid (red) line, and TDSE calculation (green) dash.
${\rm Im} f(k_z,\Delta)$ given by
\Eref{ampf}, (blue) short dash and TDSE (magenta) dots.
XUV photon energy  $\omega=106$~eV.
Delays $\Delta$ (top to bottom) are $\Delta=0.4T$, 
 $\Delta=0.5T$,  $\Delta=0.7T$,
$T$ is an optical cycle of the XUV pulse.
Vertical solid line corresponds to the momentum $k_z$ for which
$E_0+\omega=k_z^2/2$.}
\label{fig5}
\end{center}
\end{figure}

\begin{figure}[h]
  \begin{center}
    \begin{tabular}{cc}
      \resizebox{60mm}{!}{\includegraphics{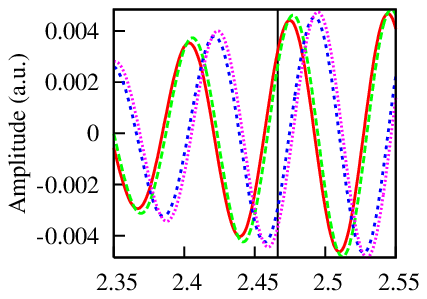}} \\
      \resizebox{60mm}{!}{\includegraphics{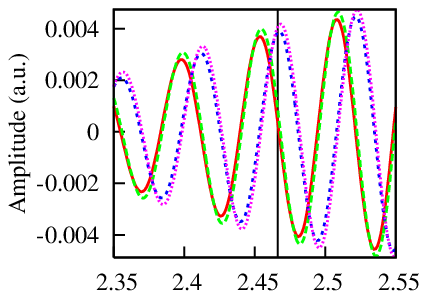}} \\
      \resizebox{60mm}{!}{\includegraphics{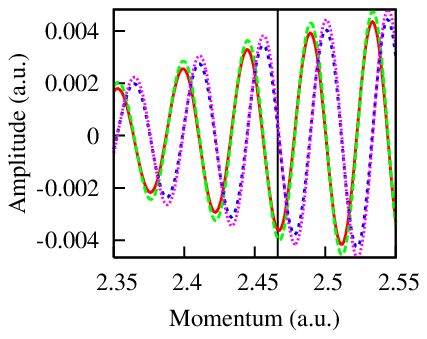}}
    \end{tabular}
\caption{(Color online) Ionization from the $2p$ state of Ne atom.
${\rm Re} f(k_z,\Delta)$ computed
using \Eref{ampf}, solid (red) line, and TDSE calculation (green) dash.
${\rm Im} f(k_z,\Delta)$ given by
\Eref{ampf}, (blue) short dash and TDSE (magenta) dots.
XUV photon energy  $\omega=106$~eV.
Delays $\Delta$ (top to bottom) are $\Delta=0.3T$, 
 $\Delta=0.4T$,  $\Delta=0.5T$,
$T$ is an optical cycle of the XUV pulse.
Vertical solid line corresponds to the momentum $k_z$ for which
$E_0+\omega=k_z^2/2$.}
\label{fig2}
\end{center}
\end{figure}

The data displayed in these figures show that
\Eref{ampf} allows to compute values of the two-color ionization
amplitude $ f(k_z,\Delta)$ with a reasonable accuracy for ionization
from $2s$ and $2p$ states of Ne provided we know the reference
amplitude $f^{\rm XUV}(k_z,0)$ as a function of the momentum.  We may
now try to solve an inverse problem of the reconstruction of the
amplitude $f^{\rm XUV}(k_z,0)$ provided that absolute values of the
two-color amplitudes $ f(k_z,\Delta)$ are known for some selected
values of the delays $\Delta$ and momenta $k_z$. This can be
demonstrated as follows.  We choose a trial form for the amplitude
$f^{\rm XUV}(k_z,0)$:
\be
f^{\rm XUV}(k_z,0)= A e^{-a(E-\epsilon)^2+i\tau(E-\epsilon)},
\label{tr}
\ee
where $E=k_z^2/2$, $a$, $\tau$, and $\epsilon$ are fitting parameters
, and $A$ is a complex number which does not depend on the energy $E$.
Parameter $\epsilon$ has a meaning of the energy at which the
cross-section of the photo-ionization by the XUV pulse is peaked.  The
first guess for the value of this parameter can be obtained from the
energy conservation $E_0+\omega=\epsilon_0$. We could fix the value of
this parameter to $\epsilon_0$. However, more accurate results are
obtained if we treat it as a fitting parameter.
The parameter $\tau$, as can be
immediately seen from the \Eref{td}, has a meaning of the time delay.

\begin{figure}[h]
  \begin{center}
      \resizebox{60mm}{!}{\includegraphics{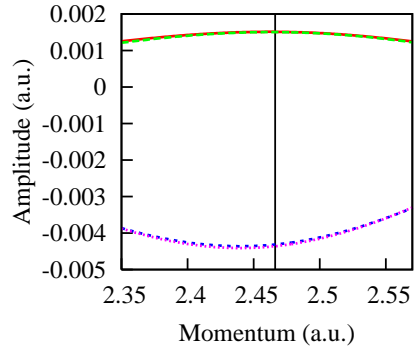}} 
\caption{(Color online) ${\rm Re} f^{\rm XUV}(k_z,0)$ computed
using \Eref{tr}, solid (red) line, and TDSE calculation (green) dash.
${\rm Im} f^{\rm XUV}(k_z,0)$ given by
\Eref{tr}, (blue) short dash and TDSE (magenta) dots.
XUV photon energy  $\omega=106$~eV, ionization from the $2p$  state of
Ne.
Vertical solid line corresponds to the momentum $k_z$ for which
$E_0+\omega=k_z^2/2$.}
\label{fig3}
\end{center}
\end{figure}

The ansatz \eref{tr} does, in fact, a very good job at reproducing the
amplitude $f^{\rm XUV}(k_z,0)$ as \Fref{fig3} testifies.  This figure
shows comparison of a fit using the functional form \eref{tr} to the
'exact' amplitude $f^{\rm XUV}(k_z,0)$ which we obtain from the TDSE
solution for ionization of the $2p$ sub-shell of the Ne atom by the XUV pulse.

By using Eqs.~\eref{tr} and \eref{ampf}, we can compute the trial 
amplitude
$f^t(k_z,\Delta)$ and, consequently, the trial
electron spectrum
$P^t(k_z,\Delta)$ of the two-colour ionization of the Ne atom as a
function of the momentum for various values of the delay $\Delta$ between
the IR and XUV pulses.
Using this distribution we can compute 
the trial expectation values of the electron momentum
$\bar k^t_z(\Delta)$ for
various $\Delta$:
\be
\bar k^t_z(\Delta)=\int P^t(k_z,\Delta) k_z\ dk_z\ ,
\label{kt}
\ee
and compare them
with the values $\bar k_z(\Delta)$ 
which we obtain from the TDSE calculation (and which can be measured
in the experiment).

We can now form a functional:
\be
D(a,\tau,\epsilon)=\sum\limits_{i} 
\left(\bar k^t_z(\Delta_i)-\bar k_z(\Delta_i)\right)^2
\label{fun}
\ee
Sum in \Eref{fun} is taken over a set of delays $\Delta$ between the
IR and XUV pulses, for which the data are available. Presently, we use
the set $\Delta=0.2T_1$, $0.3T_1$, $0.4T_1$, $0.5T_1$, and $0.7T_1$.
By minimizing the functional \eref{fun} with respect to the parameters
$a$, $\tau$, and $\epsilon$ in \Eref{tr}, we obtain the time-delay
$\tau$.

\begin{figure}[h]
  \begin{center}
      \resizebox{60mm}{!}{\includegraphics{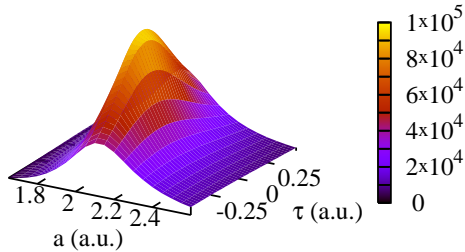}}
\caption{(Color online) Quantity $D^{-1}(a,\tau,\epsilon_0)$ as 
function of $a$, and $\tau$. Value of $\epsilon_0$ is fixed at the 
energy conservation value
$\epsilon_0=E_0+\omega$.
}
\label{fig6}
\end{center}
\end{figure}

Convergence of this procedure is illustrated in \Fref{fig6} displaying
the function $D^{-1}(a,\tau,\epsilon_0)$ for fixed value of
$\epsilon=\epsilon_0$, where $\epsilon_0=E_0+\omega$ is the value
given by the energy conservation.
This figure shows that $D^{-1}(a,\tau,\epsilon_0)$ has a well-defined
pronounced maximum in the space of the parameters $a,\tau$. This
property is very useful since it implies that
$D^{-1}(a,\tau,\epsilon)$ in \Eref{fun} has a deep minimum, which is
easy to locate.

\begin{figure}[h]
  \begin{center}
      \resizebox{60mm}{!}{\includegraphics{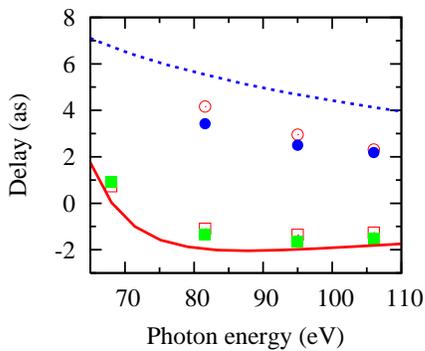}}
\caption{(Color online) Time delays computed using \Eref{td} and
the fitting procedure.
Ionization from the $2s$ state: (red) empty box- \Eref{td},
(green) solid box- the fitting procedure.
Ionization from the $2p$ state: (magenta) empty circle- \Eref{td},
(blue) solid circle- the fitting procedure. The HF results:
$2s\to Ep$ transition - (red) solid line, 
$2p\to Ed$ transition - (blue)
dashed line.
}
\label{fig4}
\end{center}
\end{figure}

Results of the procedure based on the minimization of the functional
\eref{fun} are illustrated in \Fref{fig4}, where we present the data
for the time delay for several base frequencies $\omega$ of the XUV
pulse for ionization from the $2s$ and $2p$ sub-shells of the Ne atom.
These results are compared with the values for the time delays which
we can extract directly from the TDSE calculation using the computed
amplitudes of XUV photo-ionization and \Eref{td}.

\Fref{fig4} indicates that the results of the fitting procedure described
above agree well with the results of the {\it ab initio} TDSE
calculation. For completeness, in the same figure  we display the time
delay results obtained from the Hartree-Fock elastic scattering phases
$\tau^{\rm HF} = d\delta^{\rm HF}_l/dE$.  These phases are calculated
in the frozen core Hartree-Fock approximation to electron scattering
in the field of the Ne$^+$ ion \cite{2013arXiv1302.4495K}.  The
scattering phase in the dominant photoionization channel $l = l_i+1$
is taken according to the Fano propensity rule \cite{PhysRevA.32.617},
where $l_i$ is the angular momentum of the initial bound state.
Although these results are not directly comparable to the present
calculations, which employ a localized effective potential, they
demonstrate a qualitatively similar dependence of the time delay on
the photon energy.

We may note that the present method of extraction of the attosecond
time delay, based on the soft photon approximation, can be linked to
the approach developed in \cite{pap3,clc,paz5}, which is a refinement
of the strong field approximation \Eref{str}. This refinement consists
in introducing the correction factor, multiplying the vector potential
in \Eref{str}, and adding the so-called Coulomb-laser coupling
correction in the argument of the function ${\bm A}^{\rm IR}(t)$.  It
is easy to see from Eqs.~\eref{ampf} and \eref{tr} that the
photoelectron spectrum of the two-colour ionization $P(k_z,\Delta)$ in
the soft photon approximation can be represented as
$P(k_z,\Delta)=h(k_z,\Delta+\tau)$, where $\tau$ is the time delay,
and $h$ is a periodic function of the second argument with the period
$T_1$ equal to the optical cycle of the IR field.  The first moment of
$P(k_z,\Delta)$ can, therefore, be represented as 
$\bar k_z(\Delta)=q(\Delta+\tau)$,
where $q$ is periodic with period $T_1$.  We can
write, therefore:
\be 
\bar k_z(\Delta)=\sum\limits_{n=0}^{\infty}
B_n\sin{\Omega(\Delta+\tau)}+C_n\cos{\Omega(\Delta+\tau)}
\label{ff}
\ee
The vector potential of the IR field described by  \Eref{ir} is 
$
\displaystyle A^{\rm IR}(t)={{\cal E}_0^{\rm IR}\over \Omega}
(1-\cos{\Omega t}) \ .
$
By retaining the terms with $n=0,1$ in \Eref{ff} we obtain:
\be
\bar k_z(\Delta)-b \approx cA^{\rm IR}(\Delta+\tau+\delta),
\label{ff1}
\ee
where the coefficients $b$, $c$, and $\delta$ can be expressed in terms
of the coefficients of the Fourier expansion \eref{ff}. By identifying
the coefficients $b$, $c$ and $\delta$ in th \Eref{ff1} with $k_z^0$,
the correction factor multiplying the vector potential in \Eref{str}, and
the Coulomb-laser coupling correction, we obtain the equation
replacing the strong field relation \eref{str} in the approach
developed in \cite{pap3,clc,paz5}.

\section{Conclusion}

In the present work, we examined applicability of the soft photon
approximation for evaluation of the amplitudes of two-colour XUV and IR
ionization. We used the neon atom with a model localized potential as a
convenient and representative numerical example. We
have found that the two-colour ionization amplitudes, computed using
the soft photon approximation, agree well with the {\it ab initio}
TDSE amplitudes.  This fact can be used to extract phase information
and, in particular, the time delay from the experimental photoelectron
spectra detected in attosecond streaking measurements. 
We tested the range of validity of the soft photon approximation.  We
demonstrated that this approximation renders the two-colour ionization
amplitudes accurately for the IR field intensities in the range from
$3.5\times 10^{10}$ to $5.6\times 10^{11}$~W/cm$^2$. The softness of
the IR photon requires that its frequency should be much less than the
kinetic energy of the photoelectron $\Omega/E_{\rm kin}\ll 1$.  This
means that the XUV photon energy should be well above the
photoionization threshold. This is usually the case in the attosecond
time delay measurements to minimize the effect of a large spectral
width due to a short XUV pulse.  It was found in Ref.~\cite{soft2}
that the soft photon approximation reproduces quite accurately the
angle integrated cross sections for the values of this ratio as large
as 0.06.  We observed in the present study that the amplitudes were
rendered accurately by the soft photon approximation for
$\Omega/E_{\rm kin}\approx 0.07$ for the ionization from the inner
$2s$ sub-shell of the Ne atom with the XUV frequency of 2.5~a.u. This
defines the lower bound for the XUV frequency where we can use this
approximation safely.

In our numerical examples, we confined ourselves to short XUV and IR
pulses which are used in typical attosecond streaking experiments.
However, in deriving our basic  \Eref{ampf}, we did not make
any assumptions about the pulses duration.  We can expect, therefore,
that the applicability of the soft photon approximation to the two
colour ionization process can be extended to longer IR pulses which
lead to appearance of side bands in the photoelectron spectrum.  It
can be used, therefore, for the timing analysis of the photoelectron
spectra obtained in RABITT experiments.

\section{Acknowledgment}

The authors acknowledge support of the Australian Research Council in
the form of Discovery grant DP120101805. Facilities of the National
Computational Infrastructure National Facility were used.

\np

\end{document}